\documentclass{sigchi}

\usepackage{balance}  
\usepackage{graphics} 
\usepackage{times}    
\usepackage{url}      

\usepackage{paralist}

\makeatletter
\def\url@leostyle{%
  \@ifundefined{selectfont}{\def\UrlFont{\sf}}{\def\UrlFont{\small\bf\ttfamily}}}
\makeatother
\def\pprw{8.5in}
\def\pprh{11in}

\usepackage[pdftex]{hyperref}
\hypersetup{
pdftitle={Reconstructing Browsing Activities from Browser Histories},
pdfauthor={LaTeX},
pdfkeywords={SIGCHI, proceedings, archival format},
bookmarksnumbered,
pdfstartview={FitH},
colorlinks,
citecolor=black,
filecolor=black,
linkcolor=black,
urlcolor=black,
breaklinks=true,
}

\begin{document}

\title{Reconstructing Detailed Browsing Activities\\from Browser History}

\numberofauthors{1}
\author{
  \alignauthor Geza Kovacs\\
    \affaddr{Stanford University}\\
    \email{geza@cs.stanford.edu}\\
}

\maketitle

\begin{abstract}
Users' detailed browsing activity -- such as what sites they are spending time on and for how long, and what tabs they have open and which one is focused at any given time -- is useful for a number of research and practical applications. Gathering such data, however, requires that users install and use a monitoring tool over long periods of time. In contrast, browser extensions can gain instantaneous access months of browser history data. However, the browser history is incomplete: it records only navigation events, missing important information such as time spent or tab focused. In this work, we aim to reconstruct time spent on sites with only users' browsing histories. We gathered three months of browsing history and two weeks of ground-truth detailed browsing activity from 185 participants. We developed a machine learning algorithm that predicts whether the browser window is focused and active at one second-level granularity with an F1-score of 0.84. During periods when the browser is active, the algorithm can predict which the domain the user was looking at with 76.2\% accuracy. We can use these results to reconstruct the total time spent online for each user with an $R^2$ value of 0.96, and the total time each user spent on each domain with an $R^2$ value of 0.92.
\end{abstract}


\category{H.5.m.}{Information Interfaces and Presentation (e.g. HCI)}{Miscellaneous}

\section{Introduction}

Knowing where users spend their time online, second-by-second, has numerous applications to both research and product. For example, productivity-tracking tools like RescueTime provide information about how much time users spend online on productivity and entertainment sites. Browsing activity data is also essential for studying phenomenon such as self-interruptions, where users may take a break from work to spend time on other sites.

However, gathering browsing activity data is a long and intrusive process. It requires the end user to install a monitoring application --- such as a browser extension --- that continually logs where users are spending their time, and transmits it to a server. This requires extensive permissions which may make users wary of participation, on suspicions that the extension may be malware. The user must also keep the extension installed over the duration of the study. 
Most problematically, a longitudinal study is required, with duration equivalent to the amount of browsing activity data desired.

Browsing histories, in contrast, can be instantaneously gathered by a browser extension. For a Chrome extension, this requires only a Browsing History permission, which is classified as low-risk. Browser histories can be automatically or manually scanned and filtered before sending~\cite{eyebrowse}, and can be uninstalled as soon as the history has been transmitted to the server. Most promisingly, users' browsing histories can store up to several months of historical browsing data, allowing us to instantly get results without a longitudinal study.

In this work we aim to reconstruct four pieces of browsing activity, using only browsing history:

\begin{compactitem}
	\item When is the browser focused and being actively used?
	\item What domain is the browser focused on at any given time?
	\item How much time did each user spend actively browsing?
	\item How much time did each user spend on each domain?
\end{compactitem}

These tasks are non-trivial because the browsing history represents only a thin slice: it logs events when a new page is visited, not time spent within a page or switching/closing tabs. This makes naive time-estimation heuristics fail on pages where users might spend a long time without any record in the browsing history (ie, watching a YouTube video, or scrolling down a Facebook news feed).

To train and evaluate our reconstruction mechanism, we gathered browsing histories, as well as two weeks of second-by-second browsing activities, from 185 participants recruited from Amazon Mechanical Turk. We utilize domain-related and temporal features in a random forest to outperform heuristics such as assuming a fixed time after a page visit, and are able to correctly reconstruct the time the user spent on domains with an $R^2$ value of 0.92. 

\pagebreak

\section{Related Work}

\subsection{Gathering Browsing Activities}

Gathering browsing activities by logging it in a longitudinal study is a methodology that underlies a number of studies. For example, Mark et al have conducted studies that relate browsing activities to sleep debt \cite{mark2016sleep} and stress \cite{mark2014stress}, as well as using them to investigate social media usage \cite{wang2015coming} and multitasking \cite{mark2015focused}.

Eyebrowse is an application where users can voluntarily share their browsing activities \cite{eyebrowse}. They have gathered a dataset of browsing activities from their userbase. 
We complement Eyebrowse by gathering a larger longitudinal dataset of full browsing activities and introducing a model for reconstructing attention data. If our model is successful, however, it may threaten some measure of users' security on Eyebrowse.

\subsection{Estimating User Activities from Logs}

Although no prior work has attempted to reconstruct browsing activities from browsing histories, there has been work on estimating user activities from logged data in other contexts.

Huang \emph{et al.} use mouse clicks and cursor movements to estimate users' gaze on search engine result pages \cite{huang2011no, huang2012user}. Park \emph{et al.} investigate the relationship between video view durations on Youtube and its view count, number of likes per view, and sentiment in the comments \cite{youtubeduration}. They find that these factors have significant predictive power over the duration the video, and are able to predict the duration of video views with an $R^2$ value of 0.19.

\section{Dataset}

\subsection{Dataset Collection}

\begin{figure}
    \centering
    \includegraphics[width=0.9\columnwidth]{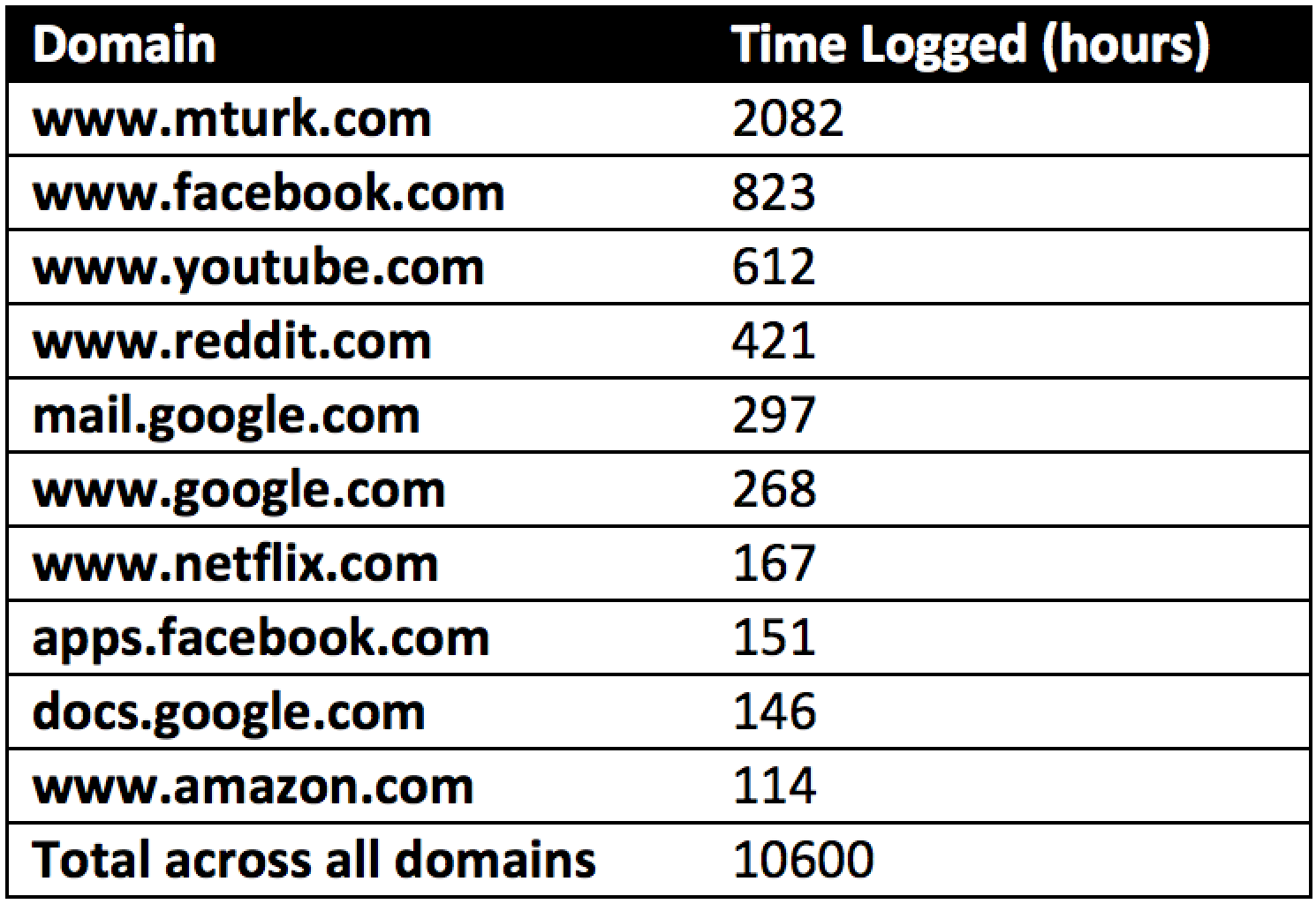}
    \caption{The top 10 domains for which we have time logged in our dataset.}
    \label{fig:top-domains}
\end{figure}

We first gathered a dataset of browsing activities and browser histories. We recruited 225 participants from Mechanical Turk, and asked them to install our extension which collects browser histories and browsing activity events (window and tab focus and switches, as well as mouse and keyboard activity such as clicks and scrolls on pages), and transmits it to our servers.

We paid users \$2 for installing the extension, and gave a bonus of \$1 for each week they kept the extension installed. We excluded users if they uninstalled the extension or became inactive for more than 3 days. We were able to gather data from 185 users in this way --- most of the 40 who dropped uninstalled the extension shortly after receiving the initial \$2 payment. We split the 185 users into training and test sets (93 users in the training set, 92 users in the test set).




In \autoref{fig:top-domains} we show the total amount of time spent on each domain, across all users. The domain with the most time spent is Mechanical Turk (as our users are from Mechanical Turk), but the other sites are all broadly used and representative of the sites used by a general audience.


\subsection{Reference Browser Activity Data}

The reference browser activity dataset was obtained by logging open, close, switch, and change events for tabs and windows via Google Chrome's tab and window APIs for extensions. We also logged when the user's screen locked or the browser became idle (defined by Chrome as 1 minute without mouse or keyboard activity), via Chrome's idle API for extensions. For each event, we logged which tabs and windows were open, the URLs they were visiting, and which tabs were focused.

We then transformed this data into spans of time, which record when the user starts and ends a period of activity on a URL: the start occurs when the URL is visited or gains tab focus, and the end occurs by navigating to a different page, closing the tab or window, switching to a different tab, browser window, or application, or if the browser becomes idle or the screen is locked.


\subsection{History Data}

The history data was obtained via Chrome's history API for extensions. It includes the URL that was visited, the time it was visited at, as well as how the visit occurred (by clicking a link, reloading a page, navigation within a frame, etc).

While attempting to find the correspondences between our history data and our reference browsing activities, we found there existed some differences. Obviously, there are many events such as tab switches and time spent scrolling down a page that are only represented in the browsing activity data. However, there are also some activities that occur in the browsing history but not the browsing activity data. One type of such event is navigation within frames, which we corrected for by eliminating them from the history (the browsing history explicitly marks navigation within frames as such). 

\section{Reconstruction Procedure}

Since our goal is to be able to reconstruct, second-by-second, whether the user's browser is active and which domain they are browsing, we broke this procedure into two parts:

\begin{compactitem}
	\item Estimate the spans during which the browser is active (as opposed to the browser being closed, idle, or a different window being in focus)
	\item Within a span in which we believe the browser to be active, estimate which domain is being viewed at each point in time.
\end{compactitem}

\pagebreak

\section{When was the browser active?}

We consider a browser to be \textit{active} at a particular second of time if the browser window is focused and there has been mouse movement/scrolling/clicking, keyboard activity, or navigation activity within the past minute. If the browser window loses focus, is closed, or the screen is locked, we consider the browser to be inactive from that second onwards.

Following common search engine practice, we consider a \textit{browsing session} to be a continuous period of time from the first second when the browser is active, until 20 minutes after the last second when the browser is active, such that there is no continuous inactive period of more than 20 minutes.

Determining when the browser is active is a classification, for each second in the browsing session, whether or not the browser is active. We consider a true positive to be when we correctly predict that the browser was active, a true negative to be when we correctly predict that the browser was inactive, a false positive to be when we predict the browser was active when it was in fact inactive, and a false negative to be when we predict the browser was inactive when it was in fact active. (We could alternatively define our task as classifying whether the browser is active or not for all seconds we have data for, including out-of-session times, but our models correctly classify all out-of-session times as being inactive, so with the exception of true negatives on out-of-session seconds, the tasks are equivalent).


If we did not have browsing history data for a user and only had aggregate data about their activities, a baseline approach might simply classify a user as being active in an all browsing sessions, or as inactive in all browsing sessions (whichever is more accurate for that particular user). This approach achieves an F1-score of 0.72 and accuracy of 0.63.


If we do have browser history, a simple model for estimating when the browser was active, is to simply guess that the browser remained active for some amount of time (for example, 1 minute or 2 minutes) after the last recorded event in the history. For example, if we set the threshold at 1 minute, this would achieve an F1-score of 0.64 and accuracy of 0.67. We tried various thresholds on our training data (each 1-minute threshold from 1 minute to 10), and found that a threshold of 5 minutes maximized both F1-score and accuracy. This model achieves an F1-score of 0.79 and accuracy of 0.76 on the test data.




We then developed a more sophisticated model for this binary classification problem using machine learning. It is based on the following intuitions:

\begin{compactitem}
	\item Browsing occurs in spans of activity -- within a continuous browsing span, the navigation activities will be densely packed.
	\item The domain may influence the expected duration of the visit -- it may be a domain that users tend to stay on for shorter or longer.
	\item The domain also influences how frequently navigation events will occur -- consider a domain that displays content in a paginated format (where navigation events will occur frequently and will be recorded in the history), versus a single-page application with infinite scrolling (where no navigation events will be recorded in the history).
\end{compactitem}

We capture the importance of the domain and the browsing spans using the following features for classification. For all time-based features, we used the logarithm of the duration.

\begin{compactitem}
    \item Time between the most recent activity and next activity in the history. If short, then the user is likely actively browsing during that entire timespan.
    \item Time since the most recent activity in the history. If short, the user is likely still on that page.
    \item Time until the next activity in the history. If short, the user may have just switched back to the browser window but has not yet made a navigation event.
    \item Domain on which the previous browsing activity occurred in the history (categorical feature with 20 categories, representing the top 20 most popular domains in the training data).
    \item Domain on which the next browsing activity occurs in the history (categorical feature with 20 categories).
    \item RescueTime productivity level of the domain (categorical feature with 5 categories, drawn from the RescueTime community).
\end{compactitem}

For the two categorical features with domains, the 20 domains we consider are the ones that had the most visits among the users in the training set. We consider only the top 20 domains because of the way categorical features are turned into a binary vector with length equal to the number of possible categories (in our case, 20 binary features to represent 20 possible domains), through a process known as one-hot encoding. To avoid the curse of dimensionality (which would lead to increased model complexity, training time, and overfitting), we consider only the top 20 domains.

Productivity levels assigned one of 5 categories to each domain: very productive, productive, neutral, distracting, or very distracting. These classifications were drawn from RescueTime, which obtained the classifications from annotations by their userbase. Domains which RescueTime does not have a productivity level for are assigned a default neutral level. As this is also a categorical feature, this is transformed into a length-5 binary feature vector via one-hot encoding.

We then train a random forest with these features, using H2O's implementation of the random forest algorithm with the default parameters \cite{randomforest}.

\begin{figure}
    \centering
    \includegraphics[width=0.9\columnwidth]{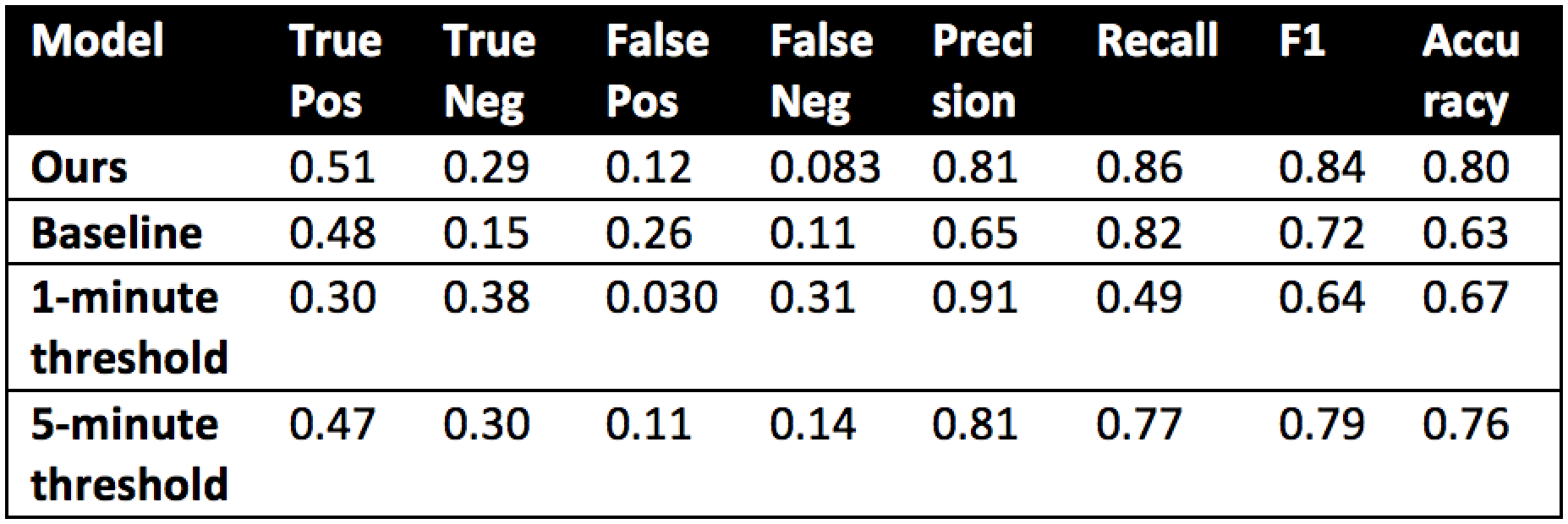}
    \caption{Performance of our machine learning method, versus various simpler approaches, on the task of predicting whether the user's browser is active at a particular second within the browsing session.}
    \label{fig:active-prediction}
\end{figure}

Our model achieves an F1 score of 0.84 and accuracy of 0.80 on the task of predicting whether the browser is active or not at a given second of time. In \autoref{fig:active-prediction}, we show the performance of our model on the task of classifying each in-session second as either active or inactive, compared to each baseline.

Our model successfully classifies all out-of-session samples as true negatives, so on the task of predicting whether the browser is active or not at all times (including out-of-session, e.g., while the user is sleeping), the precision, recall, and F1 scores remain equal, while accuracy rises to 0.96.

\section{Total time each user spent browsing}

\begin{figure}
    \centering
    \includegraphics[width=0.9\columnwidth]{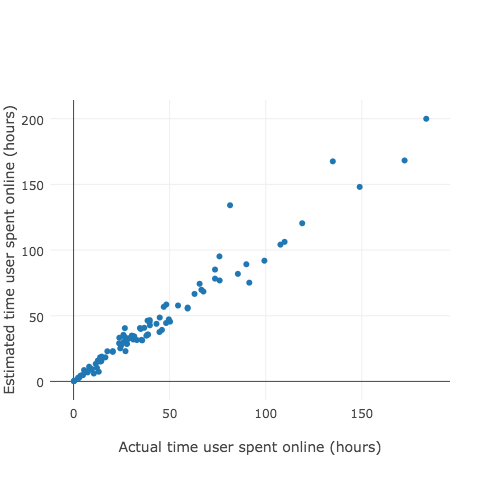}
    \caption{In this graph, for each user in our test set, we plotted a point for the total active time they spent online (x-coordinate is the actual time, and the y-coordinate is the time our algorithm estimates).}
    \label{fig:domain-timespent-plot}
\end{figure}

Now that we have reconstructed whether a user is actively using the browser at any given point in time, we can estimate the total amount of time each user spent online. In \autoref{fig:domain-timespent-plot}, for each user we have plotted the reference time the user spent online, against the time our algorithm estimates that the user spent online (reconstructed by summing the active seconds predicted by our classifier). The result is well-correlated, with an $R^2$ value of 0.96.

If we consider for each user the absolute error normalized by the total reference time spent online, and take the mean across users, the mean normalized absolute error for our predicted total online times is 0.15 ($\sigma$=0.14). If we had instead used the 5-minute-threshold classifier for determining what the active-browsing times were for each user, the mean normalized absolute error for our predicted online times would be 0.19 ($\sigma$=0.21).







\section{Which domain was focused?}

Now that we have an estimation of when the browser was active, we can determine which domain was focused at any given point in time. The active domain often does not match the most recent navigation event in the history, because users switch tabs or keep multiple windows open.

If we had only aggregate data about users' browsing activity, we might simply always predict that that the user is on the domain that they spend the most time on. This approach predicts the domain correctly on 31.6\% of seconds (among the seconds the browser is active, on the users in the test set).

If we have browsing history data for a user, a simple heuristic for predicting which domain the user is on is to assume that we are browsing the page that was visited most recently. This is able to predict the domain correctly on 74.2\% of seconds in the dataset (among the seconds the browser is active, on the users in the test set).

We developed a more sophisticated model which treats problem as a multi-class classification problem. Our model attempts to decide between 4 classes for the domain the user is currently on:

\begin{compactitem}
    \item The domain in the most recent navigation event in the history, which we will refer to as C (for ``current'')
    \item The domain in the next navigation event in the history, which we will refer to as N
    \item The domain before C in the history (not matching C), which we will refer to as P1 (``past, one back'')
    \item The domain before P1 the history (not matching C or P1), which we will refer to as P2 (``past, two back'')
\end{compactitem}

The intuition behind our model is that if a user has tabbed over to a different tab, it must have been opened at some point in the past -- this is what P1 and P2 are designed to keep track of (they approximate potential tabs that might be open in the background). The type of domain also matters -- users are more likely to keep certain common sites, such as Facebook or Gmail, open in the background than other pages. Finally, if a user has switched to a different tab, they may eventually navigate to another page from it, which will appear in the browsing history.

We choose these 4 classes because they account for most of the domains the user is on during browsing -- in our test data, 92.4\% of active browsing time will be on one of these domains. (As for the remaining 7.6\% of time, for 6.3\% the domain visit appears further back in the browsing history, while 1.3\% do not appear in the history at all -- we will discuss reasons for this in the Discussion section).

Note that these classes can overlap (ie, N can equal C, P1, or P2). In these cases, for the purpose of labeling samples in our training data, we labeled it as the most common class it could belong to (where the order of commonness is C, N, P1, P2). In the 7.6\% of seconds where the active domain did not match any of C, N, P1, or P2, we did not include the sample in our training data.

The features we used are described below. For all time-based features, we used the logarithm of the duration. We will use the shorthand t(C) to refer to the time of the most recent navigation event in the history, t(N) to refer to the time the next navigation event, t(P1) to refer to the time of the most recent history event where P2 appears, and t(P2) to refer to the time of the most recent history event where P2 appears.

\begin{compactitem}
    \item Time between t(C) and t(N). If short, the user will likely not be tabbing to other locations.
    \item Time that has elapsed since t(C). If short, the user is likely still on domain C.
    \item Time until t(N). If short and N was already open as a tab, the user may have switched to domain N.
    \item Time that has elapsed since t(P1). If long, the user is less likely to be on P1.
    \item Time that has elapsed since t(P2).
    \item \# of visits in the history that have occured since t(P1). If large, the user is less likely to be on P1.
    \item \# of visits in the history that have occured since t(P2).
    \item \# of times in the past 20 minutes that a history event on N follows an event on a different domain. If high, N is likely to be a site that the user leaves open in the background.
    \item \# of times in the past 20 minutes that a history event on C follows an event on a different domain.
    \item \# of times in the past 20 minutes that a history event on P1 follows an event on a different domain.
    \item \# of times in the past 20 minutes that a history event on P2 follows an event on a different domain.
    \item Whether the referring visit id (the source page for the navigation event) of N equals the visit id of C. If true, the user had likely stayed on C prior to opening N.
    \item Whether the referring visit id of N equals the visit id of P1. If true, the user had likely switched tabs to P1 prior to opening N.
    \item Whether the referring visit id of N equals the visit id of P2.
    \item Which domain C, N, P1, and P2 are (each is a categorical feature with 20 categories)
    \item Whether N is the same domain as C, P1, or P2 (these 3 binary features help resolve overlap in classes)
\end{compactitem}

The categorical features representing domains represent the 20 most common domains in the training set. The referring visit id feature makes use of a metadata field accessible via Chrome's history API which tells us which prior link a particular visit came from, if it was accessed by clicking a link.


We then train a random forest with these features, using H2O's implementation of the random forest algorithm with the default parameters \cite{randomforest}.

\begin{figure}
    \centering
    \includegraphics[width=0.9\columnwidth]{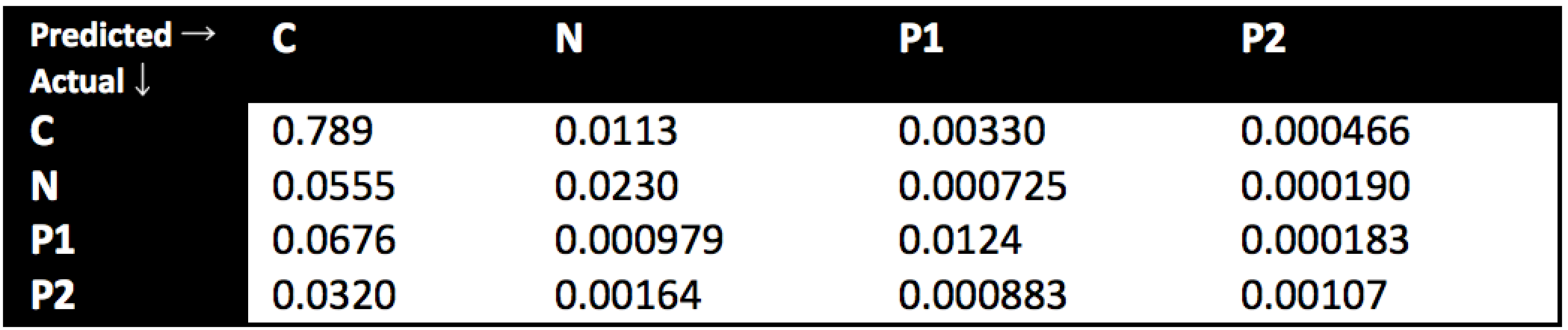}
    \caption{Confusion matrix for our random forest, which classifies each active second of browsing as either the domain seen most recently in the history (C), the next domain in the history (N), the domain before C in the history (P1), or the domain before P1 in the history (P2).}
    \label{fig:confusion-matrix}
\end{figure}

Our model correctly predicts the domain in 82.5\% of seconds where it is one of C, N, P1, or P2. The confusion matrix is shown in \autoref{fig:confusion-matrix}, showing that most errors are with rarer classes being mispredicted as more common classes. However, because in 7.6\% of the seconds the domain is not one of C, N, P1, or P2 and hence cannot be correctly classified by our model, then this results in our algorithm predicting the correct domain 76.2\% of the time (among the seconds the browser is active, on the users in the test set).




\begin{figure}
    \centering
    \includegraphics[width=0.9\columnwidth]{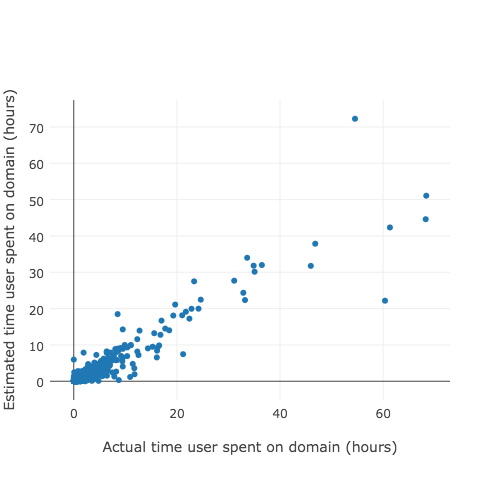}
    \caption{In this graph, for each user in our test set, we plotted a point for each domain they visited representing the time spent on the domain (x-coordinate is the actual time, and the y-coordinate is the time our algorithm estimates).}
    \label{fig:domain-timespent-plot}
\end{figure}

\section{How much time was spent on each domain?}

Having now reconstructed the domains where users spent their time on a second-by-second basis, we can now evaluate how accurately this can be used to compute overall time spent on domains. Overall time spent on domains is a useful piece of data that can be used in several time-tracking and productivity applications, as well as studies where time spent online or on a specific service is of interest.


In \autoref{fig:domain-timespent-plot}, for each user in our test set, we plotted a point for each domain they visited, representing the relation between actual time spent on the domain, versus our combined prediction model's predicted time (which was obtained by first determining the active seconds with our browser-active machine learning classifier, feeding this to our focused-domain machine learning classifier, and summing over the results). Our reconstructed total time spent on each domain is well-correlated with the actual time spent. If we take the mean of the $R^2$ value over all users, we achieve a mean $R^2$ value of 0.92 ($\sigma$=0.122). If instead of our machine learning classifiers we instead use the simpler 5-minute active-threshold and most-recent-domain heuristic classifiers, this achieves a mean $R^2$ value of 0.91 ($\sigma$=0.123).

We also computed for each user the absolute error summed over each domain prediction, and normalized it by the total time spent. With our machine learning classifiers, the mean of this normalized absolute error is 0.305 across users ($\sigma$=0.129), while with the heuristic classifiers, the mean normalized absolute error is 0.344 ($\sigma$=0.128).

\pagebreak

\section{Discussion}


We will now discuss sources of errors and limitations of our technique, and how they might be addressed.

If we look back to our plot of reconstructed total domain-focus times in \autoref{fig:domain-timespent-plot}, we see that there are a handful of outliers where we predict a much lower amount of browsing than the reference. Many of these are due to rarer video sites or long single-page articles which are not among the top 20 domains. Here, users might spend several minutes actively browsing without any record in the history. A potential way to fix this issue is to estimate the amount of time it would take a user to consume the content within a given URL, using a headless browser. For example, for video content, we could scrape the page, see if there are any videos, and detect the length of the video. We might then predict whether the user had fully watched the video or not -- based on whether the next event in the browsing history occurred around when the video would have finished playing. Analogously, for textual content, we might estimate the amount of time needed to read the article based on the amount of visible text, and develop a model to predict whether the user had fully read the article based on the surrounding browsing history. This information could then be used to correct our estimate of how much time the user had actually spent on the page.

Another underlying cause for some underestimates of time spent was that the user had partially cleared their browsing history -- Chrome provides an option to clear browsing history from the past hour. Although we had attempted to exclude users who cleared their browsing histories from our training and test datasets, our technique had only detected when users had cleared at least a day's worth of data (as our extension sent the histories to our servers on a daily basis). Hence, when making computations and inferences using browsing histories, we must consider the possibility that the user may have partially cleared their history.

We had mentioned that during 1.3\% of all active browsing seconds, the domain that the user was focused on did not appear anywhere in the preceding browsing history. Although part of this may have been due to users partially clearning their browsing histories, another cause was that URLs for certain non-http/https protocols are not logged in the history. Among these, the chrome://newtab page which is visited when the user opens a new tab accounts for nearly half (0.6\% of total active browsing seconds), while other chrome:// URLs such as the the bookmarks, downloads, settings, extension settings, and various extension-related pages contributed to another 0.1\% of total active browsing seconds.

An additional limitaiton of our technique is that browsing histories are not logged in incognito mode (this is Chrome's term for the private browsing mode), so we might not be able to capture all browsing activity from users who use the incognito feature. However, this limitation is also shared by using a browser extension to log data, as browser extensions are disabled in incognito mode by default.




\pagebreak

\section{Conclusion}

Browsing activity data, which tells us on a second-by-second basis whether the browser is active and which page is being viewed, is useful for many experiments and applications, but is difficult and time-consuming to gather as it requires a longitudinal study. Browsing histories, in contrast, are easy to gather -- we can access several months of browsing history data instantly simply by asking the user to install a Chrome extension -- but does not capture key details, such as when the browser is in focus, when the user is actively browsing a page, and when the user switches windows or tabs. 

In this paper we used browser histories to reconstruct estimates of 4 key elements of browsing activity: what times the browser is active, which domain the user is focused on when the browser is active, total time spent online, and total time spent on each domain. We first gathered a dataset by asking Mechanical Turk users to install our extension which collects both longitudinal browsing activity data as well as browser logs. We then used this gathered dataset to by training a pair of machine learning algorithms, one of which classifies whether the browser is active or not at a given second, and another which identifies which domain is focused when the browser is active. These metrics can be used to derive how much time the user spent on each domain, as well as the total time spent online.

These reconstructed browsing activities have many applications, both for research and productivity applications. For example, in the context of a productivity or time-tracking application, we can bootstrap the process with our estimates of time spent on each domain, allowing the user to see (approximate) results immediately based on several months of data. It can also be used to develop more robost sureveys, and smarter interventions: rather than asking people to self-report how much time they spend on sites like Facebook, a survey can ask the user to install an extension that will locally compute an estimate based on the user's browsing history, and fill out the question. If we collected these reconstructed browsing activities for a pool of potential participants and stored them in a database, experiments and interventions that target particular populations -- for example, users who spend over four hours on Reddit each day -- can now much more effectively recruit participants based on their browsing activities. 

Reconstructed browsing activity data could also potentially be used to gather data and identify patterns in browsing behaviors faster and at larger scale than the small datasets we can collect via longitudinal studies. We hope we might be able to use these at-scale reconstructed browsing activities to understand patterns of behaviors such as self-interruptions during web usage, and use this to develop interventions to improve users' productivity. The ability to instantly reconstruct several months' worth of browsing activities by asking the user to install an extension would open the gates to a new class of intelligent productivity-enhancement, survey, and data mining opportunities.

\section{Extension and Code}

We have developed an open-source Chrome extension and reconstruction code that allows researchers to access an end-user's reconstructed browsing activity and total time spent per-domain from their own websites, once the user has installed our extension. The Chrome extension is available at \url{https://github.com/gkovacs/browserlog} and the reconstruction code is at \url{https://github.com/gkovacs/browsing-behavior-reconstuction-analysis}

%
%
%
%
%
\balance



\bibliographystyle{acm-sigchi}
\bibliography{reconstruct}
\end{document}